\documentclass{article}
\usepackage{spconf}
\usepackage{amsmath}
\usepackage{graphicx, graphics, theorem, times, amsfonts, amsmath, amssymb, cite}
\usepackage{tikz}
\usetikzlibrary{shapes,arrows}
\usepackage{pgfplots}
\usepackage{color}
\usepackage{hyperref}
\usepackage{multirow}
\usepackage{rotating}
\usepackage{cleveref, subcaption}
\usepackage{bm}
\input{mysymbol.sty}
\usepackage[utf8]{inputenc}
\usepackage{enumitem}

\newcommand{\EE}{{\mathbb E}}

\newcommand{\QED}{\hfill\ensuremath{\blacksquare}}

\ninept

\title{Robust graph-filter identification with graph denoising regularization}
\name{Samuel Rey and Antonio G. Marques\thanks{Work supported by the Spanish Fed. Grants KLINILYCS TEC2016-75361-R, SPGraph PID2019-105032GB-I00 and  FPU17-04520, and by the Grants F661-MAPPING-UCI and F663-AAGNCS funded
		by the Comunidad de Madrid (CAM) and 
		King Juan Carlos University
		(URJC).}}
\address{ Dept. of Signal Theory and Communications, King Juan Carlos University, Madrid, Spain}
\begin{document}
%
\maketitle
\begin{abstract}
When approaching graph signal processing tasks, graphs are usually assumed to be perfectly known.
However, in many practical applications, the observed (inferred) network is prone to perturbations which, if ignored, will hinder performance. Tailored to those setups, this paper presents a robust formulation for the problem of graph-filter identification from input-output observations. Different from existing works, our approach consists in addressing the robust identification by formulating a joint graph denoising and graph-filter identification problem. Such a problem is formulated as a non-convex optimization, suitable relaxations are proposed, and graph-stationarity assumptions are incorporated to enhance performance. Finally, numerical experiments with synthetic and real-world graphs are used to assess the proposed schemes and compare them with existing (robust) alternatives. 
\end{abstract}
\section{Introduction}\label{sec:introudction}
Data is becoming not only pervasive but also more heterogeneous and intricate. One way to handle the more complex structure present in many contemporary applications is to model the structure of the data as a graph and, then, using the graph to analyze, process, and learn from the data. That is precisely the goal of graph signal processing (GSP) \cite{EmergingFieldGSP,marques2020editorial,ortega_2018_graph,djuric2018cooperative}, which in recent years has made contributions in several tasks involving signals and filters defined over irregular domains. Those include sampling and reconstruction of graph signals \cite{SamplingOrtegaICASSP14,chen2015discrete}, graph-signal denoising \cite{chen2014signal,onuki2016graph,rey2019underparametrized}, graph-filter design \cite{SandryMouraSPG_TSP14Freq,segarra2017optimal,liu2018filter} or graph-filter identification \cite{segarra2017blind,zhu2020estimating,shafipour2018directed}, to name a few. Applications range from setups where the graph is explicit --as in communication, power and social networks \cite{kolaczyk2009book,ortega_2018_graph}-- to those where the network is implicit and must be learned from the data itself --as in gene-regulatory data or brain networks \cite{kolaczyk2009book,sporns2012book}--, motivating the development of new (GSP-based) algorithms that learn the graph \cite{mateos_2019_connecting,dong_2019_learning}. 


Since GSP is a relatively recent area of research, it is not surprising that (almost all) existing works assume that the graph topology is perfectly known, focusing on how to leverage the graph structure to process and learn from the data. Nonetheless, in many practical cases the graph contains errors, including perturbations and observation noise in setups dealing with explicit networks (e.g., link failures in a power or wireless network \cite{isufi2017filtering}) or imperfections associated with the limitations of the method used to learn the graph from the data in implicit networks (e.g., the thresholding operation employed in correlation networks \cite{kolaczyk2009book}). Equally important, it is clear that those errors, if ignored, will have a negative impact on the performance of the subsequent GSP tasks. 
%
%
%

Despite their theoretical and practical relevance, the number of works dealing with robust GSP approaches is limited \cite{ceci2020graph,miettinen2019modelling,ceci2020_semtls}. Using a small perturbation analysis, \cite{ceci2020graph} first studies how link imperfections affect the spectrum of the graph Laplacian and then proposes a Bayesian framework. The focus of \cite{miettinen2019modelling} is on postulating (graphon-based) perturbation models and analyzing how those perturbations affect graph-signal operators. Differently, \cite{ceci2020_semtls} combines structural equation models (SEMs), which can be viewed as a particularization of the GSP framework, with total least squares (TLS) for graph signal inference while also inferring the perturbations of the observed graph. The limited number of works is in part due to the fact that, while many GSP algorithms are based on spectral tools, characterizing how the errors on the graph translate to its spectrum is a challenging task \cite{segarra2015stability,ceci2020graph}.

Motivated by the previous discussion, this paper presents a (non-spectral) robust GSP approach for the problem of identifying a graph filter from input-output pairs given an imperfect (perturbed) graph. Apart from its theoretical interest, graph-filter identification has been shown to be practically relevant in, e.g., explaining the structure of real-world datasets \cite{rey2019sampling,zhu2020estimating} as well as understanding the dynamics of network diffusion processes \cite{segarra2017blind,zhu2020estimating,djuric2018cooperative}. We approach the robust estimation by recasting the problem as a graph filter identification augmented with a graph-denoising regularizer, solving the resultant optimization jointly. The proposed approach works entirely in the vertex-domain, bypassing the challenges associated with robust \emph{spectral} graph theory, and yields as a byproduct an enhanced estimate of the graph. 
Since the joint estimation leads to a non-convex optimization, we first propose suitable relaxations along with low-complexity algorithms, and then, we modify the schemes to incorporate additional graph-structure potentially present in the data. Finally, we provide numerical experiments testing the proposed schemes with synthetic and real-world graphs and comparing them with existing alternatives.

\section{Preliminaries} \label{preliminaries}
Let $\ccalG=(\ccalN,\ccalE)$ denote a (possibly directed) graph, where $\ccalN$ is the set of nodes, with cardinality $N$, and $\ccalE$ is the set of edges, with $(i,j)\in\ccalE$ if $i$ is connected to node $j$. The set $\ccalN_i:=\{\, j \, |(j,i)\in\ccalE\}$ denotes the incoming neighborhood of node $i$. For a given $\ccalG$, the adjacency matrix $\bbA \in \reals ^{N\times N}$ is sparse with non-zero elements $A_{ij}$ if and only if $(j,i)\in\ccalE$. If $\ccalG$ is unweighted, the elements $A_{ij}$ are binary; if not,  the value of $A_{ij}$ captures the strength of the link from $j$ to $i$.
The focus of this paper is not on $\ccalG$, but on modeling data as graph signals defined on the nodes of $\ccalG$. Such signals can be represented as a vector $\bbx=[x_1,\ldots,x_N]^T \in  \mathbb{R}^N$ where the $i$th entry represents the signal value at node $i$. Since the signal $\bbx$ is defined on the graph $\ccalG$, the core assumption in GSP is that the properties of $\bbx$ depend on the topology of $\ccalG$.

\vspace{.1cm}\noindent \textbf{The graph-shift operator (GSO).}
The GSO $\bbS$ is defined as an $N\times N$ matrix whose entry $S_{ij}$ can be non-zero only if $i=j$ or $(j,i)\in\ccalE$. 
Common choices for $\bbS$ are $\bbA$ and the graph Laplacian $\bbL$, which is defined as  $\bbL:=\diag(\bbA\bbone)-\bbA$ \cite{EmergingFieldGSP,djuric2018cooperative}. The GSO accounts for the topology of the graph and, at the same time, represents a linear transformation that can be computed \textit{locally}. Specifically, if $\bby=[y_1,\ldots,y_N]^T$ is defined as $\bby=\bbS\bbx$, then node $i$ can compute $y_i$ provided that it has access to the values of $x_j$ at its neighbors $j\in \ccalN_i$. We assume that $\bbS$ is diagonalizable so that there exists an $N\times N$ matrix $\bbV$ and a \textit{diagonal} matrix $\bbLambda$ such that $\bbS=\bbV\bbLambda\bbV^{-1}$. The matrix $\bbV^{-1}$ is adopted as the Graph Fourier Transform (GFT) for graph signals \cite{SandryMouraSPG_TSP14Freq}, with the $N\times 1$ vector $\tbx=\bbV^{-1}\bbx$ standing for the representation of $\bbx$ on the graph frequency domain.

\vspace{.1cm}\noindent\textbf{Graph filtering.}
Graph filters, a cornerstone piece of GSP, are linear graph-signal operators that can be written as a polynomial of the GSO $\bbS$, i.e., 
\begin{eqnarray}\label{eq:GFeq}
 \! \!\!\!\!  &\bbH \!=\!\!\sum_{k=0}^K\! h_k\bbS^k \!=\!\bbV\diag(\bbPsi\bbh)\bbV^{-1}\!\!=\!\bbV\diag(\tbh)\bbV^{-1}\!\!\!&\;\;
\end{eqnarray}
where $\bbh=[h_0,...,h_K]^T$ is the vector collecting the filter coefficients,  $\bbPsi\in\reals^{N\times K}$ with $\Psi_{ij}:=(\Lambda_{ii}^{j-1})$ is a Vandermonde matrix representing the GFT for filters, and $\tbh=\bbPsi\bbh$ is a the $N\times 1$ vector representing the frequency response of the graph filter $\bbH$ \cite{SandryMouraSPG_TSP14Freq}.
When applied to an input graph signal $\bbx$, the output of the graph filter is $\bby=\bbH\bbx =\sum_{k=0}^K h_k (\bbS^k\bbx)$, where $\bbS^k\bbx$   can be viewed as a version of $\bbx$ diffused across a $k$hop neighborhood and $h_k$ are the coefficients of the linear combination \cite{segarra2017optimal}. 

\vspace{.1cm}\noindent\textbf{Graph stationarity.}
A \textit{random} graph signal $\bbx$ with zero mean and covariance $\bbC_\bbx \!= \!\EE[\bbx \bbx^T]$ is said to be stationary in the symmetric $\bbS$ if its covariance matrix $\bbC_\bbx$ is diagonalized by $\bbV$, the eigenvectors of $\bbS$~\cite{djuric2018cooperative}. Equivalently\footnote{A small technical condition must hold for these two statements to be equivalent; see~\cite{marques2017stationary}.}, a \textit{random} graph signal is defined to be stationary in $\bbS$ if $\bbC_\bbx$ can be written as a (positive-semidefinite) matrix polynomial of $\bbS$.


\vspace{.1cm}\noindent\textbf{Robust Structural Equation Models (SEMs).}
Given a graph signal $\bby$, sparse SEMs assume that the value $\bby$ at a particular node (say the $i$th one) can be explained based on: a) the values of $\bby$ at $\ccalN_i$ and b) the value of an exogenous input signal $\bbx$ at node $i$. More formally, let the $N \times M$ matrices $\bbX=[\bbx_1,...,\bbx_M]$ and $\bbY=[\bby_1,...,\bby_M]$ collect $M$ exogenous and endogenous graph signals. Then, SEMs postulate that the following relation holds $\bbY=\bbA\bbY+\bbB\bbX$,
with $A_{ii}=0$ and $\bbB$ being a diagonal matrix whose $i$th entry $B_{ii}$ accounts for the influence of $X_{im}$ on $Y_{im}$. 
Recently, \cite{ceci2020_semtls} combined SEMs with TLS (TLS-SEMs) to account for possible perturbations on the graph topology. With $\bar{\bbA}$ denoting the perturbed adjacency matrix, $\bbDelta = \bar{\bbA} - \bbA $ denoting the perturbations, and setting $\bbB=\bbI$, the TLS-SEMs approach postulates that the graphs signals in $\bbY$ satisfy 
\begin{equation}\label{eq:tls_sem_model}
    \bbY=(\bar{\bbA}-\bbDelta)\bbY+\bbX = (\bbI-\bar{\bbA}+\bbDelta)^{-1}\bbX.
\end{equation}
Leveraging \eqref{eq:tls_sem_model} and assuming that only noisy observations of $\bbY$ (denoted as $\bar{\bbY}$) are available, TLS-SEMs address the robust estimation of $\bbA$ and $\bbY$ by formulating a joint optimization over $\bbY$ and $\bbDelta$, enforcing \eqref{eq:tls_sem_model} and promoting sparsity on $\bbDelta$  \cite{ceci2020_semtls}. From the point of view of this paper, it is important to remark that while SEMs are popular in the non-GSP (statistics) literature, the (unperturbed version of the) problem in \eqref{eq:tls_sem_model} can be reformulated using GSP tools by setting $\bbS=\bbA$ and defining the filter $\bbH_{SEM}=(\bbI - \bbS)^{-1}$, so that $\bbY=\bbH_{SEM}\bbX$.

%
%

\section{Robust graph-filter identification}\label{graph_filter_id}
In this section, we present our ``robust graph-filter identification with graph denoising regularization (RFI)'' approach. 
In contrast to existing approaches that try to recast the filter identification problem in the spectral domain, we address the design in the vertex domain and leverage the fact that graph filters are matrices that commute with the GSO. As will be clear next, formulating the problem in the vertex domain not only leads to more tractable formulations but also provides a natural way to account for the imperfections on the graph. 
%
%
To be mathematically precise, suppose that $M$ pairs of graph input-output signal pairs are available and use $\bbX=[\bbx_1,...,\bbx_M]$ to denote the $N \times M$ matrix collecting the inputs and $\bbY=[\bby_1,...,\bby_M]$ the one collecting the outputs. We further assume that the $m$th output $\bby_m$ is related to the $m$th input $\bbx_m$ via the graph filter represented by the $N \times N$ matrix $\bbH$, which is a polynomial of the true GSO $\bbS$. Generative models of the form $\bby=\bbH \bbx$ (with, e.g., the filter $\bbH$ being bandlimited, the inputs being white, or the inputs being sparse) have been shown to account accurately for several types of network diffusion processes as well as a number of real-world datasets \cite{djuric2018cooperative}. With this notation at hand, we are ready to formulate our recovery approach. 


Given $\bar{\bbS}$, a perturbed version of the true GSO $\bbS$, and the (possibly noisy) input-output signals in $\bbX$ and  $\bbY$, we aim to estimate the graph filter $\bbH$ that best explains the observations under the assumption that the graph perturbations are small. 
This is achieved by solving the RFI problem
%
%
%
\begin{equation}\label{eq:general_filter_id}
\min_{\bbS\in \ccalS, \bbH}  \|\bbY\!-\!\bbH\bbX\|_F^2 + \lambda d(\bbS,\bar{\bbS}) + \beta \|\bbS\|_0 \;\;
\mathrm{\;\;s. \;t. } \;\; \bbS\bbH=\bbH\bbS, \;\; 
\end{equation}
where $\ccalS$ is the set of admissible GSOs (e.g., Laplacian matrices with zero-row sum, or adjacency matrices with no self-loops); and $d(\cdot ,\cdot )$ is a distance function that measures the similarity between $\bbS$ and $\bar{\bbS}$ and must be chosen based on prior knowledge on the nature of the graph perturbations.
The first term in the objective accounts for noise in the observations and/or modeling inaccuracies, while the $\ell_o$ norm in the third term accounts for the fact that the true GSO is sparse. Finally, the constrain $\bbS\bbH=\bbH\bbS$ captures the fact that $\bbH$ is a polynomial of the true $\hbS$. The main novelties of our approach are twofold. 
\begin{itemize}[leftmargin=1mm]
    \item First, while most works formulate the filter recovery in the spectral domain, we put forth a vertex-domain formulation. Suppose for now that the GSO is free of errors. Since $\bbH$ has the form in \eqref{eq:GFeq}, a natural formulation of the graph identification problem is to find the $\bbh=[h_0,...,h_K]^T$ minimizing $\|\bbY-(\sum_{k=0}^Kh_k\bbS^k)\bbX\|_F^2$ or, equivalently, that minimizing $\|\bbY-\bbV \diag(\bbPsi \bbh)\bbV^{-1}\bbX\|_F^2$. However, those formulations involve high-order polynomials (either of matrix $\bbS$ or of the entries of $\bbLambda$ present in $\bbPsi$) that oftentimes give rise to numerically unstable problems \cite{djuric2018cooperative}. As a result, the graph-identification task is typically reformulated in the spectral domain as finding the $\tbh$ minimizing $\|\bbY-\bbV \diag(\tbh)\bbV^{-1}\bbX\|_F^2$. Differently, thanks to the constraint $\bbS\bbH=\bbH\bbS$ in \eqref{eq:general_filter_id}, we bypass the need to compute $\bbV$ without facing the numerical problems of classical vertex-based approaches.
    \item Second, rather than considering the graph filter as the only optimization variable, we also optimize over the actual graph. From a practical point of view, there are many applications where denoising the graph is as useful as identifying the generative graph filter. From a technical point of view, considering the true graph for spectral-based approaches is challenging. It can be rigorously shown that small perturbations in $\bbS$ can lead to significant perturbations in $\bbV^{-1}$ \cite{segarra2015stability,ceci2020graph} and, even when not large, characterizing how those perturbations translate to the eigenvectors $\bbV$ and incorporating that into the optimization is not an easy task. The consideration of the constraint $\bbS\bbH=\bbH\bbS$ circumvents these problems and opens the door to introduce prior knowledge of the nature of the perturbations by tailoring the selection of the matrix distance $d(\cdot ,\cdot )$. 
\end{itemize}

\vspace{.1cm}\noindent\textbf{Modeling graph perturbations.}
A worth discussing topic is the postulation and analysis of graph-perturbation models that combine practical relevance with analytical tractability \cite{miettinen2019modelling}. Unfortunately, space limitations prevent us from engaging in that discussion and we limit ourselves to describe one model, which will motivate the formulation in the next section. With $\bbDelta=\bar{\bbS}-\bbS$  denoting the perturbation matrix, we focus on perturbations that create and destroy links. Consider first the case of unweighted GSOs. It then holds that the entries of $\bbDelta$ can be either $0$ (when no perturbation is introduced), $1$ when a link is created, and $-1$ when a link is destroyed. A simple approach is to consider that perturbations follow an independent Bernoulli distribution with the creation and destruction probabilities being equal. Alternatively, the probabilities can vary across links or, more generically, the perturbation can be rendered dependent across links, either using a multivariate correlated Bernoulli distribution or an Ising model \cite{dai2013multivariate}.  When the GSOs are weighed, the previous model can be still used, provided that it is augmented with mechanisms that set the analog value of the perturbation.



\subsection{Algorithmic approaches}

Setting $d(\bbS,\bar{\bbS})$ in \eqref{eq:general_filter_id} to $\|\bbS-\bar{\bbS}\|_0$, so that perturbations are infrequent and independent across links, yields 
%
%
\begin{equation}\label{eq:nonconvex_filter_id}
\!\min_{\bbS\in \ccalS, \bbH} \|\bbY\!-\!\bbH\bbX\|_F^2 \!+\!  \lambda \|\bbS-\bar{\bbS}\|_0 \!+\! \beta \|\bbS\|_0 \;\;\mathrm{\;\;s. \;t. } \; \bbS\bbH=\bbH\bbS, \;\; 
\end{equation}
which is a non-convex problem due to the presence of the  $\ell_0$ norms and the bilinear constraint. A simple approach to deal with the $\ell_0$  norm is to replace it with its $\ell_1$ convex counterpart (iterative re-weighted formulations are also possible). Regarding the constraint $\bbS\bbH=\bbH\bbS$, we relax it and rewrite it as a regularizer, which is better suited to an alternating minimization approach.The relaxed version of \eqref{eq:nonconvex_filter_id} can then be written as   
\begin{alignat}{2}\label{eq:convex_filter_id}
\!&\!\min_{\bbS\in \ccalS, \bbH} && \|\!\bbY\!\!-\!\!\bbH\bbX\!\|_F^2 \!+\!  \lambda \|\!\bbS\!-\!\bar{\bbS}\!\|_1 \!+\! \beta \|\bbS\!\|_1 \!+\! \gamma \|\bbS\bbH \!\!-\!\! \bbH\bbS\|_F^2 \;\; 
\end{alignat}
While still non-convex due to presence of the bilinear terms $\bbH\bbS$ and $\bbS\bbH$, the problem in \eqref{eq:convex_filter_id} is amenable to an (efficient) alternating optimization approach \cite{bezdek2003convergence} where we iterate between two steps:

\vspace{.5mm}
\noindent \textbf{Step 1: Filter Identification.} Given $\hbS$, the current estimate of the GSO, we substitute $\bbS=\hbS$ into \eqref{eq:convex_filter_id} and solve \eqref{eq:convex_filter_id} with respect to $\bbH$. This yields
\begin{alignat}{2}\label{eq:step1_filterid}
\!\!&\hbH =  \arg \min_{\bbH} && \|\bbY\!-\!\bbH\bbX\|_F^2 \!+\! \gamma \|\hbS\bbH \!-\! \bbH\hbS\|_F^2,
\end{alignat}
which is a least-squares problem whose closed-form solution is 
\begin{align}
 \mathrm{vec}(\hbH)\!=\!\big(\bbX\bbX^T\!\!\otimes\!\bbI \!+\! \gamma (\hbS\hbS^T \!\!\otimes\! \bbI \!+\! \bbI \!\otimes\! \hbS^T\hbS \!-\! \hbS^T\!\!\otimes\!\hbS^T \!-\! \hbS\!\otimes\!\hbS)\big)^{-1}\nonumber\\
 \times (\bbX\otimes \bbI)\mathrm{vec}(\bbY),\hspace{4.9cm}\nonumber
\end{align} 
where $\otimes$ is the Kroneker product and $\bbI$ has size $N \times N$.

\vspace{.5mm}
\noindent \textbf{Step 2: Graph Denoising.} Given $\hbH$, the current estimate of the filter, we substitute $\bbH=\hbH$ into \eqref{eq:convex_filter_id} and solve \eqref{eq:convex_filter_id} with respect to $\bbS$. This yields
\begin{alignat}{2}\label{eq:convex_graph_denosing}
\!\!\!\hbS \!=\!  \arg \min_{\bbS\in \ccalS}   \lambda \|\bbS-\bar{\bbS}\|_1 \!+\! \beta \|\bbS\|_1 \!+\! \gamma \|\bbS\hbH \!-\! \hbH\bbS\|_F^2  
\end{alignat}
which, provided that $\ccalS$ is convex (e.g., $\ccalS$ being the set of symmetric adjacency matrices with $A_{ii}=0$), can be handled using efficient variants of the lasso algorithm \cite{van2011sparse}. 

\begin{figure}
    \centering
    \includegraphics[width=0.32\textwidth]{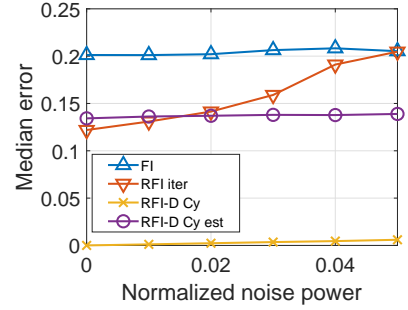}
    \vspace{-.25cm}
    \caption{Estimation performance for ``Test case 1''. Normalized median error of $\hbH$ for different levels of noise in the observed signals. ``FI'' stands for the filter-identification algorithm ignoring graph perturbations, ``RFI iter'' is the two-step iterative algorithm solving \eqref{eq:convex_filter_id}, and ``RFI-D'' is a low-complexity approximation to \eqref{eq:convex_filter_id_stationarity}. }
    \vspace{-.5cm}
    \label{fig:noise_inf}
\end{figure}

\begin{figure*}[!t]

	\centering
	\begin{subfigure}{0.32\textwidth}
		\centering
		    \includegraphics[width=1\textwidth]{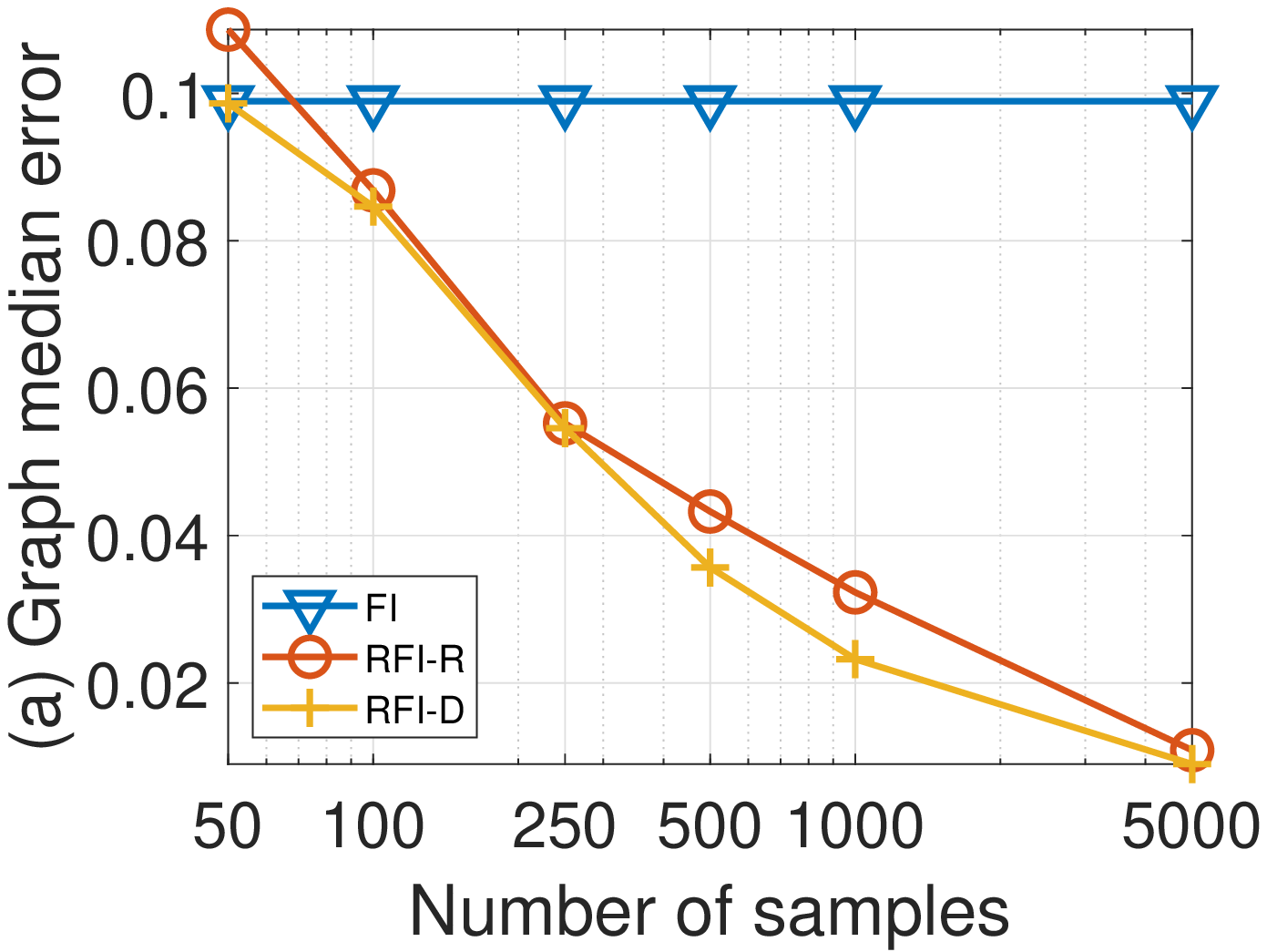}
	\end{subfigure}
	\begin{subfigure}{0.32\textwidth}
		\centering
			\includegraphics[width=1\textwidth]{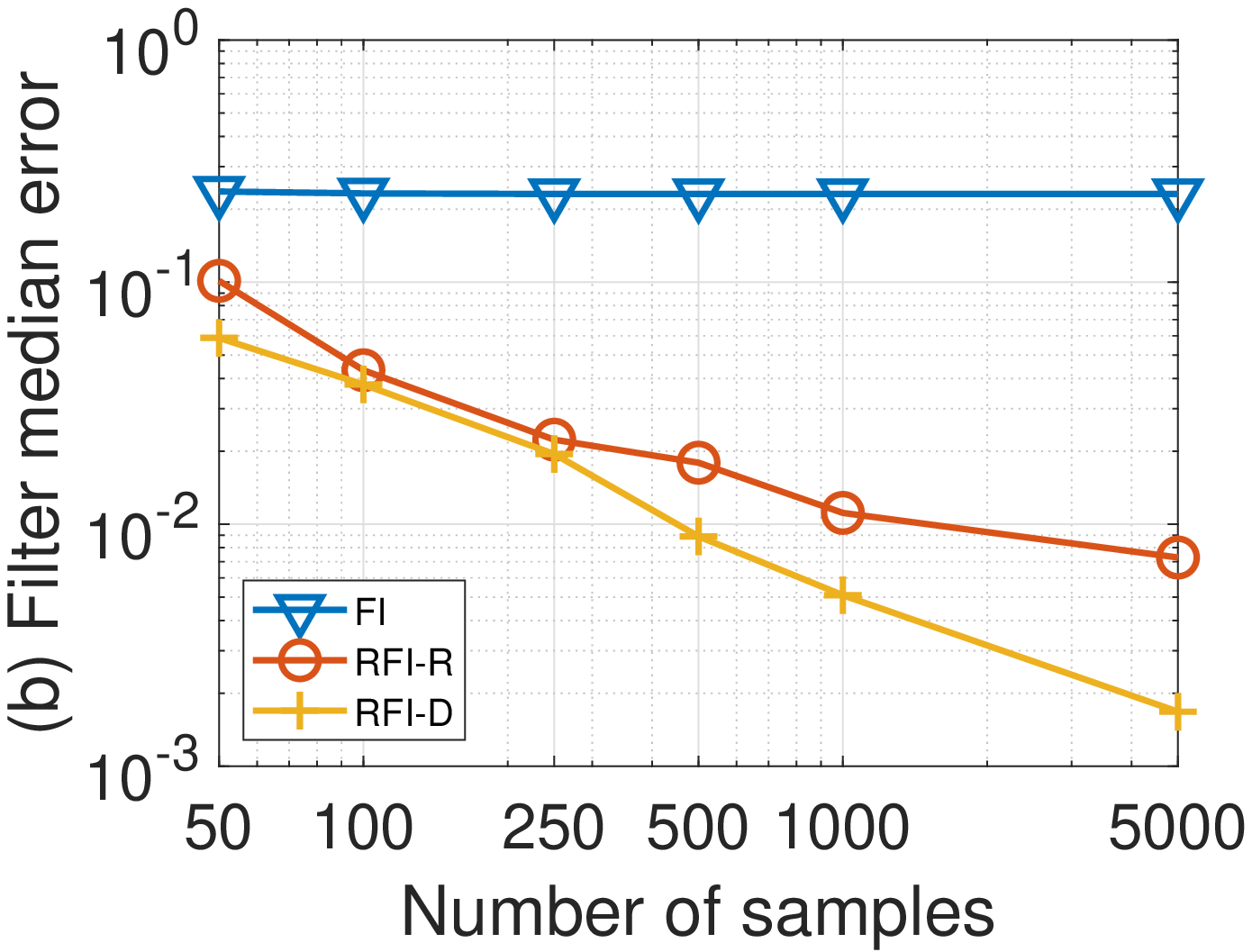}
	\end{subfigure}
	\begin{subfigure}{0.32\textwidth}
		\centering
			\includegraphics[width=1\textwidth]{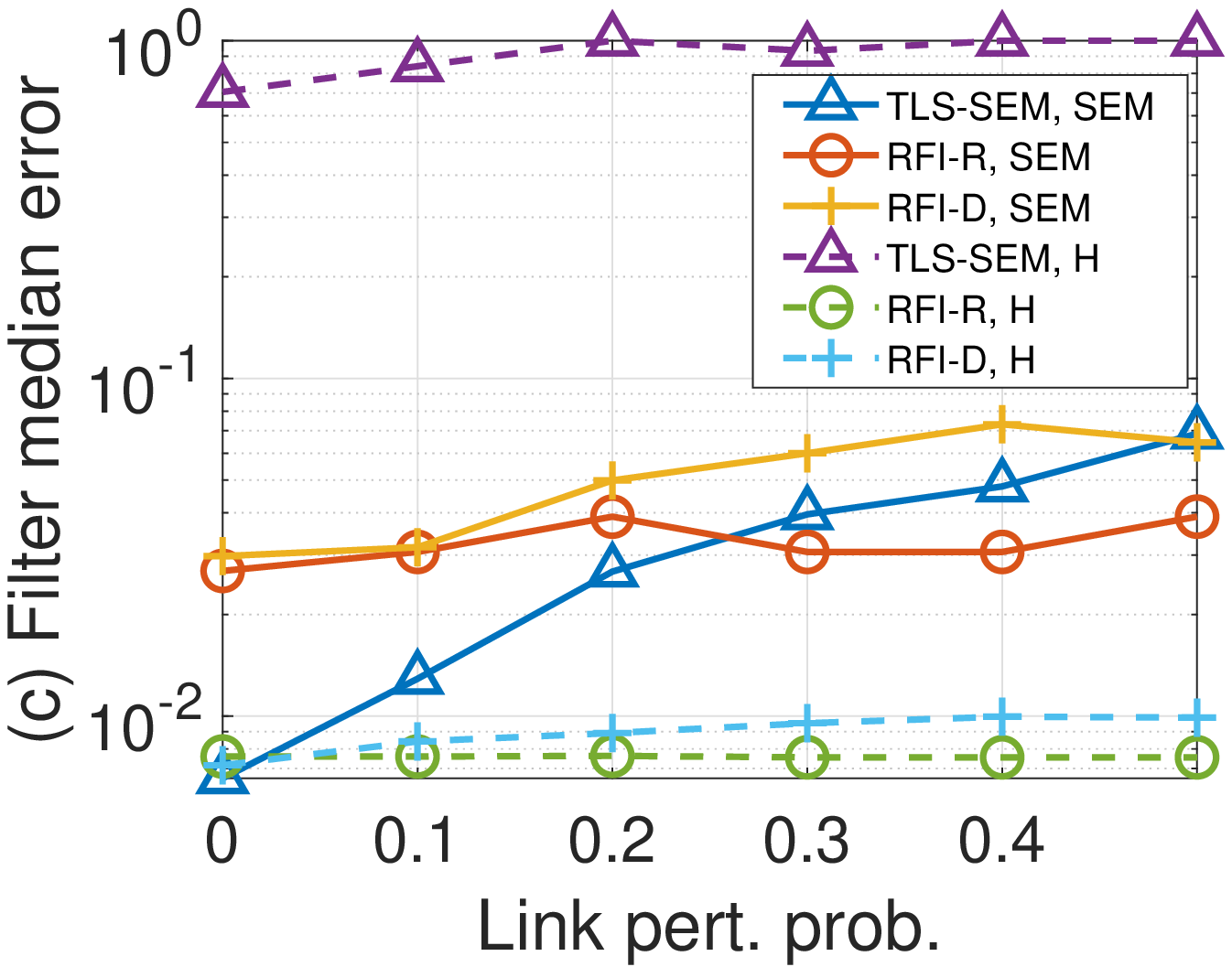}
	\end{subfigure}
	\vspace{-.25cm}
	\caption{Median estimation error for ``Test cases 2 and 3''. (a) Error estimating $\hbH$ when using the sample covariance with the Zachary's Karate graph as the number of samples $M$ grows; (b) Error denoising the GSO using the sample covariance with the Zachary's Karate graph as $M$ grows; (c) Error estimating $\hbH$ when increasing the link perturbation probability for ER graphs and two different signal generation models.
	} \label{fig:main_fig}
	\vspace{-.35cm}
\end{figure*}


\vspace{.5mm}
The alternating algorithm is initialized with $\hbS=\bar{\bbS}$ and then iterations between steps 1 and 2 are run until convergence (to a local optimum) is obtained \cite{bezdek2003convergence}. The number of iterations required to converge will be sensitive to the value of $\gamma$. When $\gamma$ is very close to zero, the two problems decouple and the solution converges quickly to that of two separate problems [cf. \eqref{eq:step1_filterid} and \eqref{eq:convex_graph_denosing} with $\gamma=0$]. If $\gamma$ is too large, the filter $\hbH$ obtained in the first iteration will be an (almost exact) polynomial of $\bar{\bbS}$ so that the algorithm will converge in one iteration to the same solution as that of the (non-robust) design that assumes that the true GSO is $\bar{\bbS}$ [cf. \eqref{eq:nonconvex_filter_id} with $\bbS=\bar{\bbS}$]. Hence, $\gamma$ must be tuned so that the algorithm can explore a larger set of solutions. In this context, adoption of schemes that start with a small $\gamma$ (encouraging exploration during the warm-up phase) and then increase $\gamma$ as the iteration index grows (guaranteeing that the final $\hbH$ is a polynomial of $\hbS$) is a suitable alternative for the setup at hand.

\subsection{Additional structure: Stationary observations}
Up to this point, we have used the fact that $\bbY$ and $\bbX$ are related via a polynomial of the GSO. However, the input/output signals can exhibit additional properties that depend on the supporting graph. Notable examples include signals being graph-bandlimited \cite{EmergingFieldGSP}, diffused sparse graph signals \cite{segarra2017blind,rey2019sampling} or graph stationary. If that is indeed the case, this additional information can be incorporated into the optimization, opening the door to an enhanced recovery performance. Due to space limitations, we focus our discussion on the case where the observations are stationary in the symmetric GSO $\bbS$. Since the covariance matrix of a graph stationarity signal can be written as a polynomial of the GSO  (cf. Section \ref{preliminaries}), we update \eqref{eq:convex_filter_id} as        
\vspace{-.05cm}
\begin{alignat}{2}\label{eq:convex_filter_id_stationarity}
\!\!&\!\min_{\bbS\in \ccalS, \bbH} && \|\!\bbY\!\!-\!\!\bbH\bbX\!\|_F^2 \!+\!  \lambda \|\!\bbS\!-\!\bar{\bbS}\!\|_1 \!+\! \beta \|\bbS\!\|_1 \!+\! \gamma \|\bbS\bbH \!\!-\!\! \bbH\bbS\|_F^2  \nonumber\\
&\;\mathrm{s. t. } && \|\bbC_Y\!\bbS\!-\!\bbS\bbC_Y\!\|_F\!\leq\! \epsilon_Y,\,\!\|\bbC_X\!\bbS\!-\!\bbS\bbC_X\!\|_F\!\leq\!\epsilon_X,  \!\!\!\!
\end{alignat}
\vspace{-.05cm}
with $\bbC_Y$ being $\bbC_X$ the (exact or estimated) covariances of $\bbY$ and $\bbX$, respectively. If the covariance is perfectly known, then the corresponding $\epsilon$ can be set to zero. Differently, when the covariance is estimated from the observations the value of $\epsilon$ must be selected based on the quality of the estimator (accounting, e.g., for the number of available observations). 
The constraints in \eqref{eq:convex_filter_id_stationarity} involve the true GSO. This implies that, when an alternating optimization is adopted, such constraints must be considered in the graph-denoising step. Alternatively, since $\bbC_Y$, $\bbC_X$ and $\bbH$ are all polynomials of $\bbS$, the equalities $\bbC_Y \bbH=\bbH\bbC_Y$ and $\bbC_X \bbH=\bbH\bbC_X$ must hold as well. Hence, one could also augment the graph identification step with the corresponding constraints. The next section will numerically test several of these alternatives, assessing their recovery performance.

\vspace{-0.5mm}
\section{Numerical experiments}\label{results}
\vspace{-0.5mm}
This section provides numerical results illustrating the estimation performance of the proposed algorithms and comparing them with TLS-SEM. Unless otherwise stated: i) the random Erdös Rényi (ER) graphs used for the synthetic-graph examples have $N=20$ nodes and edge probability $p=0.25$; and ii) perturbations on the GSO, which are independent across links, occur with a probability of 0.1.
The code, along with additional experiments, can be found in the repository {\footnotesize \url{https://git.io/JTGQH}}.

\vspace{.1cm}\noindent \textbf{Test case 1.} 
We generate 100 ER graphs. For each graph we generate a graph filter with random coefficients drawn from a uniform distribution in the interval $[-1, 1]$, and for each $(\bbS,\bbH)$ pair, $M=10$ observed signals following the graph-filter generative model assumed in Section \ref{graph_filter_id} are considered.
The inputs $\bbX$ are generated as zero mean white Gaussian signals. The random vector $\bbh$ is scaled to have unit norm and length $K=4$. Signal observations are assumed to be corrupted by additive white Gaussian noise (AWGN).  Fig. \ref{fig:noise_inf} plots the median normalized error of the estimated graph filters as the power of the observation noise increases.
``FI'' stands for the filter identification algorithm that ignores graph perturbations, equivalent to \eqref{eq:nonconvex_filter_id} with $\bbS=\bar{\bbS}$;
``RFI iter'' is the two-step iterative algorithm proposed to solve \eqref{eq:convex_filter_id}; and
``RFI-D'' implements \eqref{eq:convex_filter_id_stationarity} by first estimating $\hbS$ with $\gamma=0$ [cf. \eqref{eq:step1_filterid}] and then $\hbH$ with a value of $\gamma$ high enough to guarantee that $\hbH$ and $\hbS$ commute [cf. \eqref{eq:convex_graph_denosing}], forcing the algorithm to converge in only two steps. Thus, the ``RFI-D'' algorithm can be equivalently viewed as first denoising the GSO and then using it for identifying the graph filter.
The difference between the two versions of this algorithm is that one assumes perfect knowledge of the covariance matrix $\bbC_Y$ while the other one only has access to the (noisy) $\bbY$ and, hence, $\bbC_Y$ must be replaced with its sample estimate.
The results illustrate the benefits of the robust schemes, which outperform the non-robust alternative. Moreover, the more information is leveraged, the better the algorithm performs. As expected, the algorithm ``RFI iter'', which does not take advantage of the (sample) covariance and tries to find the exact balance between the observation and the perturbation errors, is more sensitive to noise in the observations.  

\vspace{.1cm}\noindent \textbf{Test case 2.} In this case, the performance of the proposed algorithms is evaluated on the Zachary's karate graph \cite{girvan2002community}. 
The base graph is always the same and $100$ graph filters with perturbed GSOs are considered.
The signals are generated as in the previous experiment, but we assume that the ensemble covariance is always unknown, so that the algorithms exploiting stationarity are run using the sample covariance $\hbC_Y$. The AWGN corrupting the observations has a normalized power of $0.1$. All the remaining parameters are set as in ``Test case 1''. Fig. \ref{fig:main_fig}(a) shows the median normalized error of $\hbH$  vs. the number of observations $M$. Since in this case $M$ is larger than $N$, we consider a new low-complexity robust algorithm to approximate \eqref{eq:convex_filter_id_stationarity}, referred to as ``RFI-R''. Assuming that $\hbC_Y$ is full rank and close to the actual covariance, the ``RFI-R'' algorithm replaces $\bbS\bbH-\bbH\bbS$ with $\hbC_Y\bbH-\bbH\hbC_Y$. The elimination of the bilinear terms renders \eqref{eq:convex_filter_id_stationarity} convex and separable in $\bbS$ and $\bbH$, decreasing the complexity and converging in two steps. 

To gain insights, we start by representing $\|\hbS-\bbS\|_1/(N(N-1))$, the error between the estimated and the actual graph, for the different algorithms and values of $M$; see Fig. \ref{fig:main_fig}.(a). The results confirm that, as $M$ grows, the robust methods are able to find a more accurate $\hbS$. Regarding ``FI'', where no graph denoising is implemented, the normalized error coincides with the link perturbation probability. This enhanced graph-estimation performance is expected to help in the filter identification task, as confirmed in Fig. \ref{fig:main_fig}.(b). The results reveal that as $M$ increases, the joint influence of the enhanced estimation of $\hbS$ and the larger number of observations promote an improved estimation of $\hbH$. Moreover, the two low-complexity schemes that leverage stationarity achieve good results. As expected, ``RFI-D'', which estimates $\bbH$ relying on both $\hbS$ and $\hbC_Y$, outperforms ``RFI-R'', which only relies on $\hbC_Y$. 


\vspace{.1cm}\noindent \textbf{Test case 3.} Using 100 ER graphs with $200$ observed signals each and setting the normalized power of the noise corrupting the observations to $0.1$, this test case compares the estimation performance of the proposed algorithms with that of TLS-SEMs as the probability of perturbing a link increases; see Fig. \ref{fig:main_fig}(c). Two sets of input-output signals are considered: one generated using the SEMs in \eqref{eq:tls_sem_model} (suffix ``SEM'' in the legend) and another one generated as in the previous experiments (suffix ``H''). For the cases where the signals follow the SEMs, ``TLS-SEM'' outperforms our algorithms when the link perturbation probability is small (this is expected because our algorithms need to learn the particular form of the generating filter), but the results get similar to those of ``RFI-R'' and ``RFI-D'' as this probability increases.
Differently, when data follows the model ``H''  the algorithms put forth in this paper outperform ``TLS-SEM''. This was expected due to mismatch between the simpler model assumed by ``TLS-SEM'' and the data. In contrast, the good performance of the proposed solutions on both set of signals illustrates the benefits of considering more general modeling assumptions.



\newpage

\vfill\pagebreak
\bibliographystyle{IEEE}
\bibliography{citations}

\begin{thebibliography}{10}

\bibitem{EmergingFieldGSP}
D.I. Shuman, S.K. Narang, P.~Frossard, A.~Ortega, and P.~Vandergheynst,
\newblock ``The emerging field of signal processing on graphs: Extending
  high-dimensional data analysis to networks and other irregular domains,''
\newblock {\em IEEE Signal Process. Mag.}, vol. 30, no. 3, pp. 83--98, May
  2013.

\bibitem{marques2020editorial}
A.~G. Marques, N.~Kiyavash, J.~M.~F. Moura, D.~Van~De Ville, and R.~Willett,
\newblock ``Graph signal processing: Foundations and emerging directions
  (editorial),''
\newblock {\em IEEE Signal Process. Mag.}, vol. 37, Nov. 2020.

\bibitem{ortega_2018_graph}
A.~{Ortega}, P.~{Frossard}, J.~{Kovacevic}, J.~M.~F. {Moura}, and
  P.~{Vandergheynst},
\newblock ``Graph signal processing: Overview, challenges, and applications,''
\newblock {\em Proc. IEEE}, vol. 106, no. 5, pp. 808--828, 2018.

\bibitem{djuric2018cooperative}
P.~Djuric and C.~Richard,
\newblock {\em Cooperative and Graph Signal Processing: Principles and
  Applications},
\newblock Academic Press, 2018.

\bibitem{SamplingOrtegaICASSP14}
A.~Anis, A.~Gadde, and A.~Ortega,
\newblock ``Towards a sampling theorem for signals on arbitrary graphs,''
\newblock in {\em IEEE Intl. Conf. Acoust., Speech and Signal Process.
  (ICASSP)}, 2014, pp. 3864--3868.

\bibitem{chen2015discrete}
S.~{Chen}, R.~{Varma}, A.~{Sandryhaila}, and J.~{Kovačević},
\newblock ``Discrete signal processing on graphs: Sampling theory,''
\newblock {\em IEEE Trans. Signal Process.}, vol. 63, no. 24, pp. 6510--6523,
  2015.

\bibitem{chen2014signal}
S.~Chen, A.~Sandryhaila, J.M.F. Moura, and J.~Kovacevic,
\newblock ``Signal denoising on graphs via graph filtering,''
\newblock in {\em IEEE Global Conf. Signal and Info. Process. (GlobalSIP)}.
  IEEE, 2014, pp. 872--876.

\bibitem{onuki2016graph}
M.~Onuki, S.~Ono, M.~Yamagishi, and Y.~Tanaka,
\newblock ``Graph signal denoising via trilateral filter on graph spectral
  domain,''
\newblock {\em IEEE Trans. Signal Info. Process. Networks}, vol. 2, no. 2, pp.
  137--148, 2016.

\bibitem{rey2019underparametrized}
S.~Rey, A.~G. Marques, and S.~Segarra,
\newblock ``An underparametrized deep decoder architecture for graph signals,''
\newblock in {\em IEEE Intl. Wrksp. Computat. Advances Multi-Sensor Adaptive
  Process. (CAMSAP)}. IEEE, 2019, pp. 231--235.

\bibitem{SandryMouraSPG_TSP14Freq}
A.~Sandryhaila and J.M.F. Moura,
\newblock ``Discrete signal processing on graphs: Frequency analysis,''
\newblock {\em IEEE Trans. Signal Process.}, vol. 62, no. 12, pp. 3042--3054,
  June 2014.

\bibitem{segarra2017optimal}
S.~Segarra, A.~G. Marques, and A.~Ribeiro,
\newblock ``Optimal graph-filter design and applications to distributed linear
  network operators,''
\newblock {\em IEEE Trans. Signal Process.}, vol. 65, no. 15, pp. 4117--4131,
  2017.

\bibitem{liu2018filter}
J.~Liu, E.~Isufi, and G.~Leus,
\newblock ``Filter design for autoregressive moving average graph filters,''
\newblock {\em IEEE Trans. Signal Info. Process. Networks}, vol. 5, no. 1, pp.
  47--60, 2018.

\bibitem{segarra2017blind}
S.~Segarra, G.~Mateos, A.~G. Marques, and A.~Ribeiro,
\newblock ``Blind identification of graph filters,''
\newblock {\em IEEE Trans. Signal Process.}, vol. 65, no. 5, pp. 1146--1159,
  2017.

\bibitem{zhu2020estimating}
Y.~Zhu, F.~J. Iglesias-Garcia, A.~G. Marques, and S.~Segarra,
\newblock ``Estimating network processes via blind identification of multiple
  graph filters,''
\newblock {\em IEEE Trans. Signal Process.}, vol. 68, pp. 3049--3063, 2020.

\bibitem{shafipour2018directed}
R.~{Shafipour}, S.~{Segarra}, A.~G. {Marques}, and G.~{Mateos},
\newblock ``Directed network topology inference via graph filter
  identification,''
\newblock in {\em IEEE Data Science Wrksp. (DSW)}, 2018, pp. 210--214.

\bibitem{kolaczyk2009book}
E.~D. Kolaczyk,
\newblock {\em Statistical Analysis of Network Data: Methods and Models},
\newblock Springer, New York, NY, 2009.

\bibitem{sporns2012book}
O.~Sporns,
\newblock {\em Discovering the Human Connectome},
\newblock MIT Press, Boston, MA, 2012.

\bibitem{mateos_2019_connecting}
G.~{Mateos}, S.~{Segarra}, A.~G. {Marques}, and A.~{Ribeiro},
\newblock ``Connecting the dots: Identifying network structure via graph signal
  processing,''
\newblock {\em IEEE Signal Process. Mag.}, vol. 36, no. 3, pp. 16--43, May
  2019.

\bibitem{dong_2019_learning}
X.~{Dong}, D.~{Thanou}, M.~{Rabbat}, and P.~{Frossard},
\newblock ``Learning graphs from data: A signal representation perspective,''
\newblock {\em IEEE Signal Process. Mag.}, vol. 36, no. 3, pp. 44--63, 2019.

\bibitem{isufi2017filtering}
E.~Isufi, A.~Loukas, A.~Simonetto, and G.~Leus,
\newblock ``Filtering random graph processes over random time-varying graphs,''
\newblock {\em IEEE Trans. Signal Process.}, vol. 65, no. 16, pp. 4406--4421,
  2017.

\bibitem{ceci2020graph}
E.~Ceci and S.~Barbarossa,
\newblock ``Graph signal processing in the presence of topology
  uncertainties,''
\newblock {\em IEEE Transactions on Signal Processing}, vol. 68, pp.
  1558--1573, 2020.

\bibitem{miettinen2019modelling}
J.~Miettinen, S.~A. Vorobyov, and E.~Ollila,
\newblock ``Modelling graph errors: Towards robust graph signal processing,''
\newblock {\em arXiv preprint arXiv:1903.08398}, 2019.

\bibitem{ceci2020_semtls}
E.~Ceci, Y.~Shen, G.~B. Giannakis, and S.~Barbarossa,
\newblock ``Graph-based learning under perturbations via total least-squares,''
\newblock {\em IEEE Trans. Signal Process.}, vol. 68, pp. 2870--2882, 2020.

\bibitem{segarra2015stability}
S.~Segarra and A.~Ribeiro,
\newblock ``Stability and continuity of centrality measures in weighted
  graphs,''
\newblock {\em IEEE Trans. Signal Process.}, vol. 64, no. 3, pp. 543--555,
  2015.

\bibitem{rey2019sampling}
S.~Rey, F.~J. Iglesias, C.~Cabrera, and A.~G Marques,
\newblock ``Sampling and reconstruction of diffused sparse graph signals from
  successive local aggregations,''
\newblock {\em IEEE Signal Process. Lett.}, vol. 26, no. 8, pp. 1142--1146,
  2019.

\bibitem{marques2017stationary}
A.~G. Marques, S.~Segarra, G.~Leus, and A.~Ribeiro,
\newblock ``Stationary graph processes and spectral estimation,''
\newblock {\em IEEE Trans. Signal Process.}, vol. 65, no. 22, pp. 5911--5926,
  2017.

\bibitem{dai2013multivariate}
B.~Dai, S.~Ding, and G.~Wahba,
\newblock ``Multivariate {B}ernoulli distribution,''
\newblock {\em Bernoulli}, vol. 19, no. 4, pp. 1465--1483, 2013.

\bibitem{bezdek2003convergence}
J.~C. Bezdek and R.~J. Hathaway,
\newblock ``Convergence of alternating optimization,''
\newblock {\em Neural, Parallel \& Scientific Computations}, vol. 11, no. 4,
  pp. 351--368, 2003.

\bibitem{van2011sparse}
E.~Van~den Berg and M.~P. Friedlander,
\newblock ``Sparse optimization with least-squares constraints,''
\newblock {\em SIAM J. Optimization}, vol. 21, no. 4, pp. 1201--1229, 2011.

\bibitem{girvan2002community}
M.~Girvan and M.~E.~J. Newman,
\newblock ``Community structure in social and biological networks,''
\newblock {\em Proc. of the national academy of sciences}, vol. 99, no. 12, pp.
  7821--7826, 2002.

\end{thebibliography}

\end{document}